# Tetragonal to orthorhombic phase transition in SmFeAsO: a synchrotron powder diffraction investigation

A. Martinelli[1,*], A. Palenzona[1,2], C. Ferdeghini[1], M. Putti[1,3], E. Emerich[4]

[1] Artificial and Innovative Materials Laboratory, National Institute for the Physics of Matter – National Research Council, corso Perrone 24, 16152 Genova - Italy
[2] Dept. of Chemistry and Industrial Chemistry, University of Genoa, via Dodecaneso 31, 16146 Genova - Italy
[3] Dept. of Physics, University of Genoa, via Dodecaneso 33, 16146 Genova – Italy
[4] SNBL at ESRF, 6 rue Jules Horrowitz, 38043 Grenoble – France

* corresponding author: amartin@chimica.unige.it

**Abstract**
The crystal structure of SmFeAsO has been investigated by means of Rietveld refinement of high resolution synchrotron powder diffraction data collected at 300 K and 100 K. The compound crystallizes in the tetragonal *P*4/*nmm* space group at 300 K and in the orthorhombic *Cmma* space group at 100 K; attempts to refine the low temperature data in the monoclinic *P*112/*n* space group diverged. On the basis of both resistive and magnetic analyses the tetragonal to orthorhombic phase transition can be located at $T \sim 140$ K.

**1. Introduction**
The recent discovery of relatively high superconducting transition temperature ($T_C$) in LaFeAs(O,F) [1] attracted much attention on the *RE*FeAsO (*RE* = rare earth) class of compounds. These phases crystallizes in the tetragonal system at room temperature and their structure is built up by two kinds of planar layers constituted of edge sharing tetrahedra stacked along the *c*-axis. The former layer is constituted of tetrahedra centred by O with the *RE* at vertices (charge reservoir layer), whereas in the latter Fe coordinates As (conducting layer).

The F-free compounds exhibit a magnetic transition around 140-150 K [1,2,3,4,5,6,7] due to the development of a commensurate antiferromagnetic (AFM) spin density wave (SDW) with a small moment in the FeAs plane, as revealed by neutron diffraction analysis [3]. The arising of this magnetic structure has been related to a structural transition; in LaFeAsO this phase transition has been investigated by means of both neutron [3] and synchrotron powder diffraction analysis [8]. In the former case an abrupt structural distortion was observed at ~155 K, evidenced by the splitting of the 220 diffraction peak, and the new structure was found to belong to the monoclinic *P*112/*n* space group. A subsequent synchrotron powder diffraction analysis led to the conclusion that the low temperature phase crystallizes in the orthorhombic *Cmma* space group [8]. A synchrotron powder diffraction analysis of the iso-structural NdFeAsO revealed a similar phase transition [9] that authors relates to the occurrence of the orthorhombic structure, although the monoclinic one is not rule out.

In this work we report an investigation of the crystal structure of SmFeAsO at 100 K and 300 K carried out by means of Rietveld refinement of high resolution synchrotron powder diffraction data. Details concerning the preparation, the resistive and magnetic characterization of the sample are described in detail in ref [10]. High resolution synchrotron powder diffraction analysis was carried out at the European Synchrotron Radiation Facility (ESRF) in Grenoble (France) at the BM1B beamline. Data were collected at 300 K and 100 K ($\lambda = 0.50035$ Å; step = $0.008°$ 2θ; angular range = $1.0 – 40.5°$ 2θ) and structural refinement was carried out applying the Rietveld method using the program FullProf [11]. The peak shape was modelled using a pseudo-Voigt function; in the final cycle the following parameters were refined: the scale factor; the zero point of detector; the background (six parameters of the $5^{th}$ order polynomial function); the unit cell parameters; the atomic displacement parameters; the peak shape, the Lorentzian isotropic strain and asymmetry parameters; the atomic site coordinates.

The diffraction data collected at 300 K were successfully refined in the tetragonal *P*4/*nmm* space group and the obtained structural data (reported in Table 1) are in excellent agreement with those reported in ref [10], obtained on the same sample, but by means of laboratory X-ray powder diffraction analysis.

In both LaFeAsO and NdFeAsO phase transition is evidenced by the splitting of selected diffraction peaks, such as 220 and 322 [3,8,9]. In SmFeAsO no evident peak splitting can be detected comparing the diffraction data collected at 300 K and 100 K, but a closer examination of these peaks reveals an anomalous broadening towards high angles in the data collected at low temperature (Figure 1). A good refinement using the data collected at 100 K can be obtained using the tetragonal *P*4/*nmm* space group structural model but a careful analysis of the plot evidences that the afore mention diffraction peaks are not well fitted (Figure 2, inset on the left). A notable improvement is obtained adopting the *Cmma* structural model (Figure 2, inset on the right), already successfully applied in the case of LaFeAsO and NdFeAsO [8,9]. From the statistical point of view the two structural models differ essentially in the number of parameter used to describe the structure. In order to check if the improvement obtained changing from $P4/nmm$ to $Cmma$ is not merely related to the addition of a new parameter (that is $a \neq b$), the significance test was carried out on the crystallographic $R$ factors [12]. As a result the improvement in the agreement between observed and calculated structure amplitudes is significant and the orthorhombic structural model must be preferred.

Attempts to refine the structure applying the monoclinic $P112/n$ space group, as reported for LaFeAsO [3], led to a divergence of the refinement and thus this structural model can be ruled out.

As the orthorhombic structure takes place the Sm-As bond length splits with a notable decrease parallel to [100] (Table 2); conversely the remaining bond lengths as well as the geometry of both SmO and FeAs tetrahedra are not affected by the structural transition.

In conclusion SmFeAsO undergoes a tetragonal to orthorhombic phase transition on cooling; similarly to the LaFeAsO and NdFeAsO cases this structural change is likely to occur in the temperature region where anomalies are observed in both resistive and magnetic curves [3,8]. In the case of SmFeAsO it can be located at $T \sim 140$ K [10].

Table 1: Strutural data of SmFeAsO obtained after Rietveld refinement using the diffraction data collected at 300 K and 100 K.

| | | 300 K | | | | 100 K | | | |
|---|---|---|---|---|---|---|---|---|---|
| Space group | | P4/nmm | | | | Cmma | | | |
| Cell edge | a (Å) | 3.9390(1) | | | | 5.5611(1) | | | |
| | b (Å) | / | | | | 5.5732(1) | | | |
| | c (Å) | 8.4980(1) | | | | 8.4714(2) | | | |
| | | | x | y | z | | x | y | z |
| Wyckoff position | Sm | 2c | ¼ | ¼ | 0.1372(1) | 4g | 0 | ¼ | 0.1374(1) |
| | O | 2a | ¾ | ¼ | 0 | 4a | ¼ | 0 | 0 |
| | Fe | 2b | ¾ | ¼ | ½ | 4b | ¼ | 0 | ½ |
| | As | 2c | ¼ | ¼ | 0.6599(2) | 4g | 0 | ¼ | 0.6599(2) |
| R-Bragg (%) | | 7.48 | | | | 6.78 | | | |
| Rf-factor (%) | | 3.76 | | | | 3.65 | | | |

Table 2: Comparison among selected bond lengths in SmFeAsO at 300 K and 100 K.

| | 300 K | 100 K |
|---|---|---|
| Sm-O × 4 (Å) | 2.289(1) | 2.287(1) |
| Sm-As × 4 (Å) | 3.276(1) | 3.273(1) |
| | | 3.268(1) |
| Fe-As × 4 (Å) | 2.393(1) | 2.389(1) |

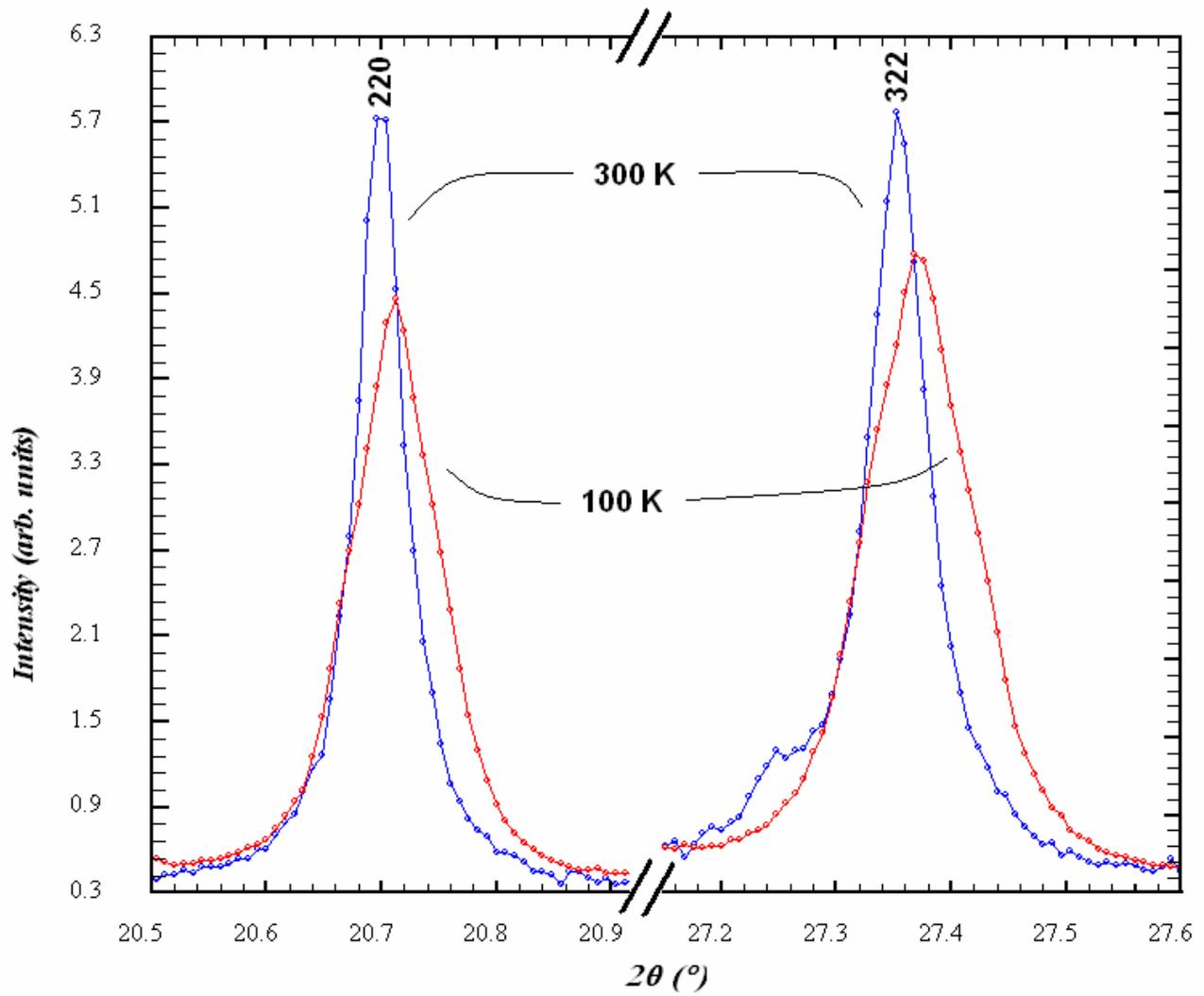

Figure 1 (colour online): Evolution of the 220 and 322 diffraction peaks (tetragonal indexing) with temperature (normalised data); anomalous broadening towards high angles can be detected in the data collected at 100 K.

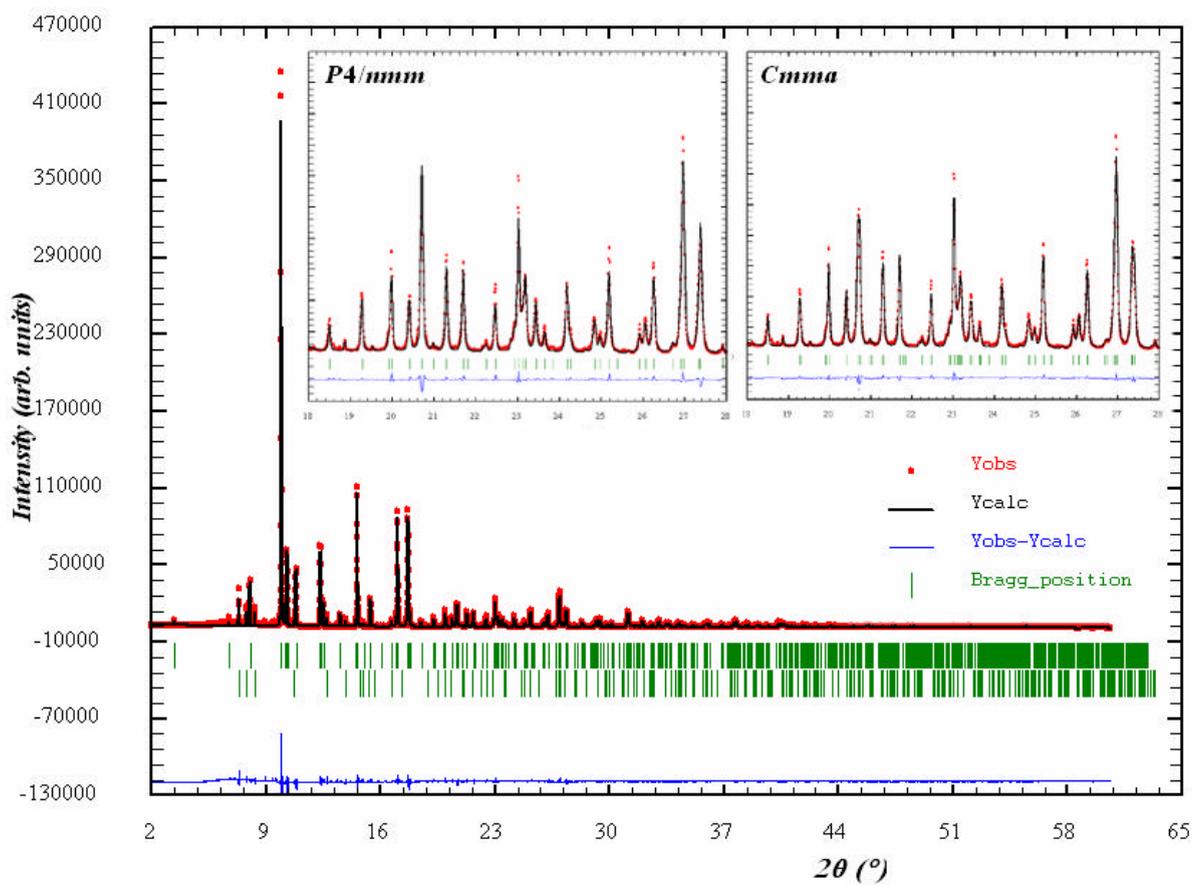

Figure 2 (colour online): Rietveld refinement plot obtained using the data collected at 100 K and applying the orthorhombic *Cmma* structural model; the insets are enlarged views of the plots obtained using the *Cmma* and *P*4/*nmm* structural models in the regions of the 220 (~20.7°) and 322 (~27.3°) diffraction peaks.